\documentclass[twocolumn,aps,showpacs,floatfix,superscriptaddress]{revtex4}
\usepackage{amsmath,amssymb,eucal,graphicx}
\begin{document}
\title{Condensates in Driven Aggregation Processes}
\author{E.~Ben-Naim}
%\email{ebn@lanl.gov} 
\affiliation{Theoretical Division and Center for Nonlinear Studies,
Los Alamos National Laboratory, Los Alamos, New Mexico 87545}
\author{P.~L.~Krapivsky} 
%\email{paulk@bu.edu}
\affiliation{Department of Physics and Center for Molecular
Cybernetics, Boston University, Boston, Massachusetts, 02215}
\begin{abstract}
  We investigate aggregation driven by mass injection. In this
  stochastic process, mass is added with constant rate $r$ and
  clusters merge at a constant total rate $1$, so that both the total
  number of clusters and the total mass steadily grow with time.
  Analytic results are presented for the three classic aggregation
  rates $K_{i,j}$ between clusters of size $i$ and $j$. When
  $K_{i,j}={\rm const}$, the cluster size distribution decays
  exponentially. When $K_{i,j}\propto i+j$ or $K_{i,j}\propto i\times
  j$, there are two phases: (i) a condensate phase with a condensate
  containing a finite fraction of the mass in the system as well as
  finite clusters, and (ii) a cluster phase with finite clusters
  only. For $K_{i,j}\propto i+j$, the cluster size distribution,
  $c_k$, has a power-law tail, $c_k\sim k^{-\gamma}$ in either phase.
  The exponent is a non-monotonic function of the injection rate:
  $\gamma=r/(r-1)$ in the condensate phase, $r<2$, and $\gamma=r$ in
  the cluster phase, $r>2$.
\end{abstract}
\maketitle
  
\section{Introduction} 

Aggregation processes in which small objects merge irreversibly to
form larger clusters are ubiquitous in nature \cite{mvs,fl}. For
example, aggregation underlies the evolution of planetary systems in
astrophysics \cite{sc}, cloud formation and dust accumulation in
atmospheric sciences \cite{pk,rld,skf,jns}, as well as polymer and gel
formation in chemical physics \cite{pjf,whs,pjf1}. Aggregation also
plays a central role in the theory of percolation \cite{sa}, fractal
formation \cite{ws}, and network growth \cite{ba,dm}.

Often, aggregation is driven by a constant injection of mass and
consequently, the total mass grows indefinitely with time
\cite{FS,white,CS,ernst,ap,crl}.  This is the case in chemical
kinetics, and in particular polymerization where the aggregation rate
depends on the number of clusters. Here, due to aggregation, the total
number of clusters typically decreases with time or saturates at a
finite value. In such aggregation processes, the system may undergo
gelation: a giant cluster develops and eventually, it contains all of
the mass in the system \cite{zhe,pvd,aal}.

In this study, we are interested in complementary aggregation
processes that describe the growth of random structures such as random
trees and random graphs, relevant in computer and information science
\cite{bb,dja,jlr,bk}.  Here, the system consists of an ensemble of
clusters where both the number of clusters and the total mass grow
with time. For example, in the Internet, clusters are autonomous
systems; injection models the creation of new autonomous systems and
aggregation describes merger of different autonomous systems
\cite{fkbcrf}.

We investigate aggregation processes with mass injection where the
total merger rate does not depend on the total number of
clusters. Our main finding is that in such situations there is
condensation rather than gelation. The system develops condensates
that contain a finite fraction of the mass. These macroscopic
condensates co-exist with microscopic clusters that contain the rest
of the mass in the system.

In our formulation, mass is injected at a constant rate and clusters
merge at a constant total rate.  We address the three classic kernels
$K_{i,j}$ for aggregation between clusters if size $i$ and $j$,
respectively: the constant rate $K_{i,j}={\rm const}$, the sum rate
$K_{i,j}=i+j$, and the product rate $K_{i,j}=i\times j$.

For the constant aggregation rate, the cluster size distribution
decays exponentially with the cluster size. For both the sum and the
product rates, where the aggregation rate grows with the aggregate
size, the system undergoes a phase transition as a function of the
injection rate. When the injection rate is smaller than some critical
value, the system is in a condensate phase. A finite fraction of the
total mass constitutes a macroscopic condensate, but the remaining
fraction of mass is in the form of microscopic clusters. The
condensate and the finite clusters coexist. When the injection rate is
larger than the critical rate, there are only finite clusters.

In the cluster phase, the distribution of cluster size generally
decays as a power-law with the cluster size. However, different
behaviors emerge in the condensate phase. For the product rate, the
size distribution falls-off exponentially at large size. For the sum
rate, however, the behavior is always power-law. Interestingly, in the
latter case, the decay exponent is not monotonic. It decreases
monotonically with the injection rate in the condensate phase but
increases monotonically in the cluster phase.

The rest of this paper is organized as follows. We introduce the
model, describe some of its basic features, and formulate the master
equation approach in Sec.~II. The constant rate, the sum rate, and the
product rate are analyzed in sections III, IV, V,
respectively. Generally, our focus is the size of the condensate and
the tail of the cluster size distribution. We conclude in Sec.~VI.
Technical derivations of two results in sections IV and V are
presented in appendices~\ref{log} and \ref{giant}.

\section{The Model}

In our model, there are two independent and competing processes: mass
injection and merger of clusters. In the first process, monodisperse
elemental clusters are added to the system. This injection process
occurs at a constant rate.  In the second process, two clusters are
merged. The mass of the resulting cluster is equal to the sum of the
two original masses.  The total merger rate is constant as well, and
since both processes occur with constant rates, we may set one of them
to unity without loss of generality. We therefore set the mass
injection rate to $r$ and the merger rate to one. Also, since the
injection is monodisperse, we set this injection size as the mass
unit. Initially, the system contains no clusters.

The total mass and the total number of clusters follow directly from
these definitions. Each injection event increases the number of
clusters by one and similarly, each merger event decreases the number
of clusters by one. Thus, the average total number of clusters,
$N(t)$, satisfies $dN/dt=r-1$ and consequently, there is simple linear
growth
\begin{equation}
\label{nt}
N(t)=(r-1)t.
\end{equation}
We restrict our attention to situations where the number of clusters
grows with time, $r>1$.  

Merger events conserve the total mass, and hence, the mass changes
only through injection events. With each injection event, unit-size
mass is added to the system, and thus, the average total mass, $M(t)$,
obeys \hbox{$dM/dt=r$}. Consequently, the total mass also grows
linearly with time,
\begin{equation}
\label{mt}
M(t)=r\,t.
\end{equation}

We investigate the cluster size distribution.  Let $C_k(t)$ be the
average number of clusters of size $k$ at time $t$. This quantity
satisfies the master equation
\begin{equation}
\label{Ck-eq}
\frac{dC_k}{dt}=r\delta_{k,1}
+\frac{1}{2}\sum_{i+j=k}K_{i,j}C_iC_j-\sum_{i}K_{i,k}C_iC_k.
\end{equation}
The initial condition is $C_k(0)=0$.  The first term accounts for mass
injection and the last two terms account for merger. The kernel
$K_{i,j}$ is defined as the aggregation rate between two clusters with
size $i$ and $j$.  The total merger rate is constant, thereby implying
the following constraint on the aggregation rate
\begin{equation}
\label{norm}
1=\frac{1}{2}\sum_{i,j}K_{i,j}C_iC_j.
\end{equation}
The total density is of course $N(t)=\sum_k C_k(t)$.  Summing the
master equation (\ref{Ck-eq}), and using the constraint (\ref{norm}),
we confirm the linear growth of the total density
(\ref{nt}). Similarly, the total mass is $M(t)=\sum_k k C_k(t)$. Only
the first term in the master equation affects the evolution of the
total mass, and summing the rate equations, we recover (\ref{mt}).

We analyze the three classic aggregation rates: the constant
aggregation rate $K_{i,j}\propto {\rm const}$, the sum aggregation
rate $K_{i,j}\propto i+j$, and the product aggregation rate
$K_{i,j}\propto i\times j$ \cite{fl}, each representing a distinct
growing random structure.  Generally, random structures such as random
trees and random graphs are made of nodes interconnected by
links. Mass injection represents addition of isolated nodes, and
merger represents addition of a link between two nodes \cite{dja,bk}.
In the constant rate process, two random structures are picked at
random and an added link connects the two. In the product rate
process, two nodes, picked at random, are connected by an added
link. Finally, the sum rate is a hybrid of the constant and the
product cases: the link connects a randomly selected node and a
randomly selected structure. We note that the constant aggregation rate
models an ensemble of random growing trees (no cycles are formed),
while the sum and the product rates model an ensemble of random
growing graphs.

\section{Constant aggregation rate}

For the constant aggregation rate, all pairs of clusters merge at the
same rate, irrespective of their size. This is the simplest and the
most widely used aggregation process with examples including fractal
aggregates \cite{pm}, domain growth \cite{ajb}, and random trees
\cite{hmm,bkm}.  We consider the following size-independent 
aggregation rate
\begin{equation}
\label{kij-const}
K_{i,j}=\frac{2}{N^2}.
\end{equation}
This constant satisfies the normalization (\ref{norm}). It decreases
as the number of clusters increases so that the overall merger rate
does not change with time.

Substituting this constant rate into the master equation
(\ref{Ck-eq}), the cluster size density satisfies 
\begin{equation}
\label{Ck-eq-const}
\frac{dC_k}{dt}=r\,\delta_{k,1}+
\frac{1}{N^2}\sum_{i+j=k}C_i\,C_j-\frac{2}{N}\,C_k.
\end{equation}
The linear growth of the total density (\ref{nt}) and the total mass
(\ref{mt}) suggest that the cluster size distribution also grows
linearly with time. Indeed, the density of the smallest clusters obeys
$dC_1/dt=r-2C_1/N$ so that this quantity, too, grows linearly with
time, $C_1(t)=N(t)\frac{r}{r+1}$.  Thus, we write the cluster size
density as a product of the overall density $N(t)$ and the
time-independent cluster size distribution $c_k$,
\begin{equation}
\label{ck-def}
C_k(t)=N(t)\,c_k.
\end{equation}
The cluster size distribution is normalized, $\sum_k c_k=1$.  This
form is consistent with the initial condition, and it satisfies
(\ref{Ck-eq-const}) when the cluster size distribution obeys the
recursion equation
\begin{equation}
\label{ck-eq-const}
(1+r)c_k=r\,\delta_{k,1}+\sum_{i+j=k}c_i\,c_j.
\end{equation}

We utilize the generating function technique \cite{gkp} to solve this
equation. Let $f(z)$ be the generating function 
\begin{equation}
\label{fz-def}
f(z)=\sum_{k=1}^\infty c_k\, z^k.
\end{equation}
Normalization implies $f(1)=1$.  Multiplying Eq.~(\ref{ck-eq-const})
by $z^k$ and summing over $k$, the generating function satisfies the
quadratic equation $f^2(z)-(1+r)f(z)+r\,z=0$.  The solution is
therefore
\begin{equation}
\label{fz-sol-const}
f(z)=\frac{r+1}{2}\left[1-\sqrt{1-\frac{4r}{(1+r)^2}z}\,\right].
\end{equation}
We can confirm that $f(z)=\frac{r}{1+r}z+\cdots$, in agreement with the
above expression for $C_1$. Writing the generating function as a
power-series, the cluster size distribution is obtained explicitly,
\begin{equation}
\label{ck-sol-const}
c_k=\frac{1+r}{4\sqrt{\pi}}\frac{\Gamma(k-1/2)}{\Gamma(k+1)}
\left[\frac{4r}{(1+r)^2}\right]^k
\end{equation}
where $\Gamma(x)$ is the Gamma function. Using the asymptotic property 
$\Gamma(x+r)/\Gamma(x)\simeq x^r$ as $x\to\infty$, we find 
\begin{equation}
\label{ck-tail-const}
c_k\simeq \alpha\,k^{-3/2}\,\beta^k
\end{equation}
with the constants $\alpha=\frac{1+r}{4\sqrt{\pi}}$ and
$\beta=\frac{4r}{1+r}$. Therefore, the cluster size distribution
decays exponentially, but there is an algebraic correction. As the
injection rate approaches the merger rate, $r\to 1$, the cluster size
distribution becomes algebraic $c_k\sim k^{-3/2}$. This limiting case
coincides with the well-known behavior for time-independent constant
aggregation rates \cite{CS}.

\section{Sum aggregation rate}

Aggregation rates proportional to the sum of the cluster sizes are
relevant in polymerization \cite{rmz}, coagulation under shear-flows
\cite{rld}, and random graphs \cite{dja}. Subject to the constraint
(\ref{norm}), the sum aggregation rate is
\begin{equation}
\label{kij-sum}
K_{i,j}=\frac{i+j}{NM}.
 \end{equation}

In this case, the master equation (\ref{Ck-eq}) becomes 
\begin{equation}
\label{Ck-eq-sum}
\frac{dC_k}{dt}=r\,\delta_{k,1}+
\frac{k}{2NM}\,\sum_{i+j=k}C_i\,C_j-\frac{kN+M}{NM}\,C_k.
\end{equation}

We again seek a solution of the form (\ref{ck-def}). The cluster size
distribution remains normalized, $\sum_k c_k=1$, and it obeys the
following recursion relation 
\begin{equation}
\label{ck-eq-sum}
(k+R)\,c_k=R\delta_{k,1}+\frac{k}{2}\sum_{i+j=k}c_i\,c_j
\end{equation}
with the constant 
\begin{equation}
\label{R-sum}
R=\frac{r^2}{r-1}.
\end{equation}
This recursion relation can be manually solved to find
$c_1=\frac{R}{R+1}$, $c_2=\frac{1}{R+2}\left(\frac{R}{R+1}\right)^2$,
etc. This procedure can be formally related to integer partitions
following the solution procedure of Ref.~\cite{bk-rods}. Such a
solution is useful only when the cluster size distribution decays
sharply. Although it is difficult to obtain an explicit analytic
solution as in (\ref{ck-sol-const}), it is still possible to obtain
many of the interesting properties of cluster size distribution from
asymptotic analysis of the generating function.

The generating function (\ref{fz-def}) obeys the nonlinear ordinary
differential equation
\begin{equation}
\label{fz-eq-sum}
z\left(f-1\right)\frac{df}{dz}=R\left(f-z\right). 
\end{equation}
Derivatives of the generating function at $z=1$ are related to the
moments of the cluster size distribution. For example, the average
cluster size follows from the first derivative, $\langle k\rangle =
f'(1)$.

Differentiating (\ref{fz-eq-sum}) and then substituting $z=1$ yields a
quadratic equation for the average cluster size
\begin{equation}
\label{kav-eq-sum}
R^{-1}\langle k\rangle^2-\langle k\rangle+1=0. 
\end{equation}
Naively, one expects that the average cluster size is the ratio
between the total mass and the total number of clusters, $\langle
k\rangle= \frac{M}{N} =\frac{r}{r-1}$. Indeed, this is a solution of
(\ref{kav-eq-sum}). However, there is another solution $\langle
k\rangle =r$.  This solution is not physical when $r>2$ since the
product $N\langle k\rangle$ can not exceed the total mass in the
system. Each solution is relevant in the appropriate range of the
parameter $r$, so that the full solution is
\begin{equation}
\label{kav-sol-sum}
\langle k\rangle =
\begin{cases}
r&                 1<r<2\\
\frac{r}{r-1}& 2<r.
\end{cases}
\end{equation}
This assertion is supported by analysis below. The total mass
contained by the clusters, $M_c=\sum_k k C_k$, is of course
proportional to the average cluster size, \hbox{$M_c=N\sum_k
kc_k=N\langle k\rangle$}. Therefore, when $r<2$, finite clusters
contain only a fraction of the mass when $r<2$, but they contain all
of the mass when $r>2$.

This behavior can be reconciled with mass conservation only if there
is a condensate of mass $M_*$ that contains the remaining fraction
$m_*=M_*/M$ of the mass in the system as follows
\begin{equation}
\label{m*-sol-sum}
m_* =
\begin{cases}
2-r  &1<r<2\\
0     &2<r.
\end{cases}
\end{equation}
Thus the system undergoes a phase transition. When $1<r<2$,
there is a condensate that contains a finite fraction of the
mass. This condensate coexists with the finite clusters that contain
the rest of the mass. The condensate contains nearly all of the mass
in the limit $r\to 1$: when injection is very slow, the condensate
``preys'' on newly added mass. As the transition point is approached,
the condensate mass vanishes, $m_*\to 0$ as $r\to 2$. When $r>2$, the
system contains only ordinary clusters.

\begin{figure}[t]
\includegraphics[width=0.4\textwidth]{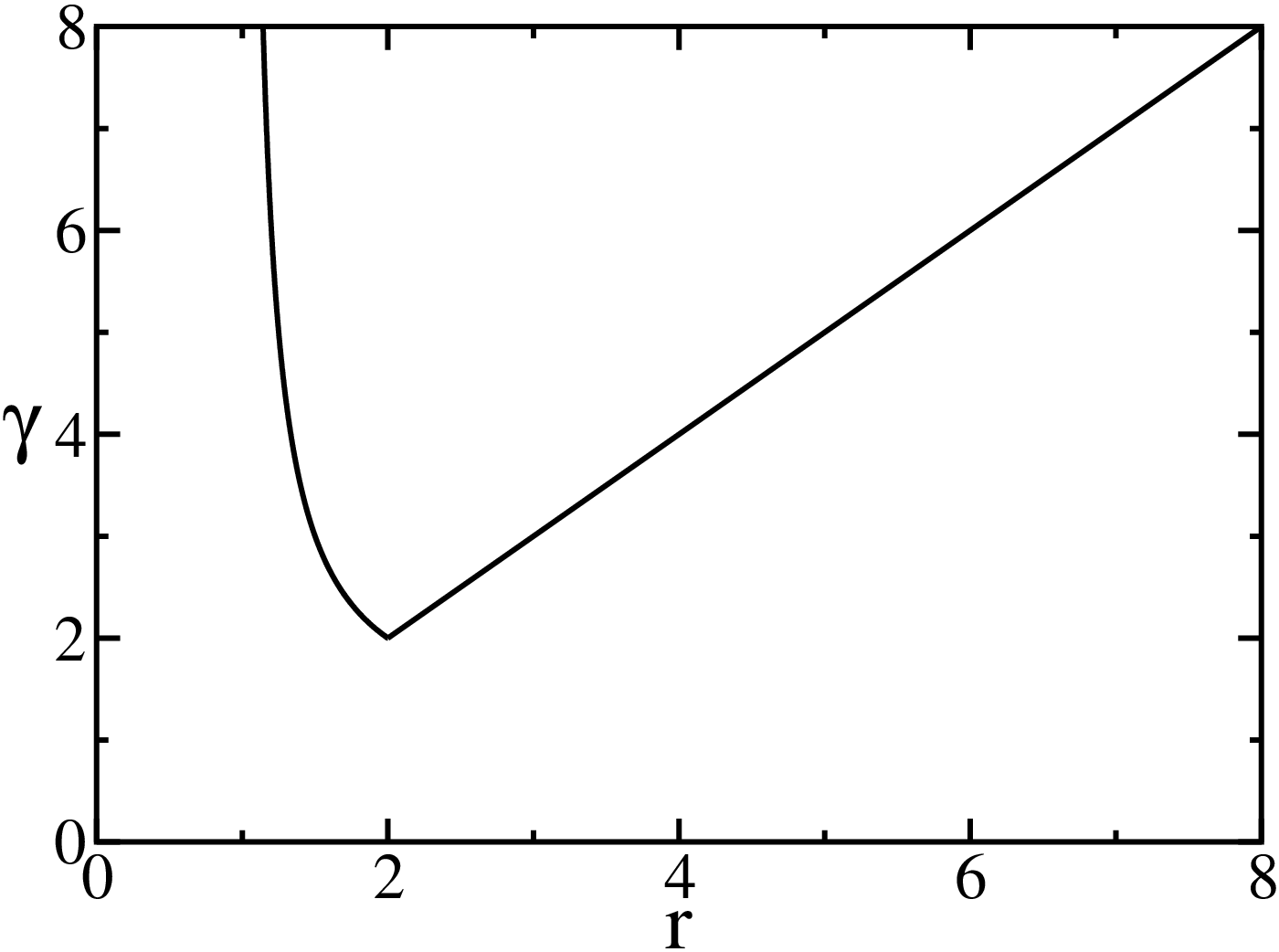}
\caption{The exponent $\gamma$ versus $r$ for the sum rate.}
\label{fig-gamma-sum}
\end{figure}

The tail of the cluster size distribution can be evaluated from the
$z\to 1$ behavior of the generating function.  In general, $f(z)$ may
contain both a regular component and a singular component,
$f(z)=f_{\rm reg}(z)+f_{\rm sing}(z)$. The regular component is a
power series in $(z-1)$.  Let us assume a singular behavior with the
leading behavior $f_{\rm sing}\propto (1-z)^{\gamma-1}$ as $z\to 1$,
\begin{equation}
\label{fz-def-sum}
f(z)=\underbrace{1+\langle k\rangle (z-1)+\cdots}_{f_{\rm reg}(z)}+
\underbrace{A(1-z)^{\gamma-1}+\cdots}_{f_{\rm sing}(z)}.
\end{equation}
In the limit $z\to 1$, this form satisfies the governing equation
(\ref{fz-eq-sum}) when 
\begin{equation}
\label{gamma-eq-sum}
\langle k\rangle \gamma=R. 
\end{equation}
The algebraic form of the generating function implies an algebraic
form for the tail of the cluster size distribution,
\begin{equation}
\label{ck-tail-sum}
c_k\sim k^{-\gamma},
\end{equation}
as $k\to \infty$. The exponent is found by substituting
(\ref{kav-sol-sum}) into the relation (\ref{gamma-eq-sum}). 
Therefore, there are two regimes of behavior 
 \begin{equation}
\label{gamma-sol-sum}
\gamma=
\begin{cases}
\frac{r}{r-1}&1<r<2\\
r&2<r.
\end{cases}
\end{equation}
We note two remarkable features. First, the cluster size density is
algebraic both in the condensate phase and in the cluster phase.
Second, the characteristic exponent is a non-monotonic function of the
injection rate: it decreases monotonically with $r$ in the condensate
phase and it increases monotonically in the cluster phase
(Fig.~\ref{fig-gamma-sum}).

The exponent $\gamma$ is minimal, $\gamma=2$, at the phase transition
point, $r=2$. For $\gamma<2$, mass conservation would be violated
because the sum $\sum_k k c_k$ is divergent.  The restriction
$\gamma>2$ justifies our previous choice of the nontrivial solution
$\langle k\rangle = r$ in Eq.~(\ref{kav-sol-sum}).

The behavior at the phase transition point requires a special
treatment. We find that the cluster size density
decays slightly faster than $k^{-2}$, namely
\begin{equation}
\label{ns-crit}
c_k\simeq \frac{2}{k^2\,(\ln k)^2}\quad {\rm as}\quad k\to\infty.
\end{equation}
The derivation of this result is detailed in Appendix A.  With this
logarithmic correction, the sum $\sum_k k c_k$ or the total mass in
finite clusters, remain finite.

In ordinary gelation (or percolation), the cluster size distribution
is algebraic {\em only} at the critical point; away from criticality,
it has exponential tails. In the present case, the cluster size
distribution exhibits a strikingly different behavior --- it is
algebraic everywhere while at the critical point there is a
logarithmic correction.

The size of the condensate can be obtained directly by focusing on its
dynamics. Let us assume that there is a giant cluster in the system
with mass $M_*$. Its growth rate is $dM_*/dt=\sum_i K_{i,M_*}\,i$ and
substituting the aggregation rate (\ref{kij-sum}), we arrive at
\begin{equation}
\label{M*-eq-sum}
\frac{dM_*}{dt}=\sum_i \frac{M_*+i}{N\,M}\,i\approx \frac{M_*(M-M_*)}{N\,M}.
\end{equation}
In the second step, we made the approximation \hbox{$M_*+i\approx
M_*$} as the condensate is much larger than the rest of the
clusters. When the condensate contains a finite fraction of the mass,
$M_*=m_*M$, then from (\ref{M*-eq-sum}), we find 
$m_*(r-1)=m_*(1-m_*)$. Solving this equation, we recover the
condensate mass (\ref{m*-sol-sum}).

This approach can also be used to derive the size of the largest
cluster in the cluster phase. From (\ref{m*-sol-sum}), the quantity
$M_*$ is negligible compared with $M$, and therefore, the governing
equation (\ref{M*-eq-sum}) reduces to $dM_*/dt\approx N^{-1}M_*\approx
\frac{1}{(r-1)t}M$ so that
\begin{equation}
\label{largest-sum}
M_*\sim t^{\frac{1}{\gamma-1}},
\end{equation}
with $\gamma=r$ when $r>2$. Therefore, the size of the largest cluster
grows algebraically with time. This result can be alternatively
derived using the algebraic behavior (\ref{ck-tail-sum}) and the
extreme statistics criterion $1\simeq N\sum_{k\geq M_*}c_k$.

\section{Product aggregation rate}

The product aggregation rate models polymerization and gelation
\cite{pjf,whs}, as well as random graphs \cite{bb,bk}, and in our case
it has the following form
\begin{equation}
\label{kij-prod}
K_{i,j}=2\frac{i\times j}{M^2}.
\end{equation}

The explicit rate equation for the product aggregation rate is
\begin{equation}
\label{Ck-eq-prod}
\frac{dC_k}{dt}=r\,\delta_{k,1}+\frac{1}{M^2}\sum_{i+j=k}i\,j\, C_iC_j
-\frac{2}{M}k\,C_k.
\end{equation}
Since the total mass appears in the denominator in the rate equation,
we use a different normalization
\begin{equation}
\label{ck-def-prod}
C_k=M\,c_k.
\end{equation}
The average number of clusters of size $k$ still
grows linearly with time. The transformation (\ref{ck-def-prod})
reduces the master equation (\ref{Ck-eq-prod}) to the nonlinear
recursion equation
\begin{equation}
\label{ck-eq-prod}
r\,c_k=r\,\delta_{k,1}+\sum_{i+j=k}i\,j\,c_i\,c_j-2k\,c_k.
\end{equation}
Given of the structure of this equation, it is convenient to use a 
different definition of the generating function
\begin{equation}
\label{fz-def-prod}
f(z)=\sum_{k=1}^\infty k\,c_k z^k.
\end{equation}
The generating equation satisfies the very same equation
(\ref{fz-eq-sum}) with 
\begin{equation}
\label{R-def-prod}
R=\frac{r}{2}.
\end{equation}
Although the governing equation is the same, the boundary condition is
different. Since the product aggregation rate grows with the cluster
size, we expect that again there are two phases: a condensate phase
and a cluster phase. The total mass contained in finite clusters is
given by $M_c=Mf(1)$. Therefore, in the condensate phase $f(1)<1$
while in the cluster phase $f(1)=1$. This change in the boundary
condition results in a drastically different behavior.

\begin{figure}[t]
\includegraphics[width=0.4\textwidth]{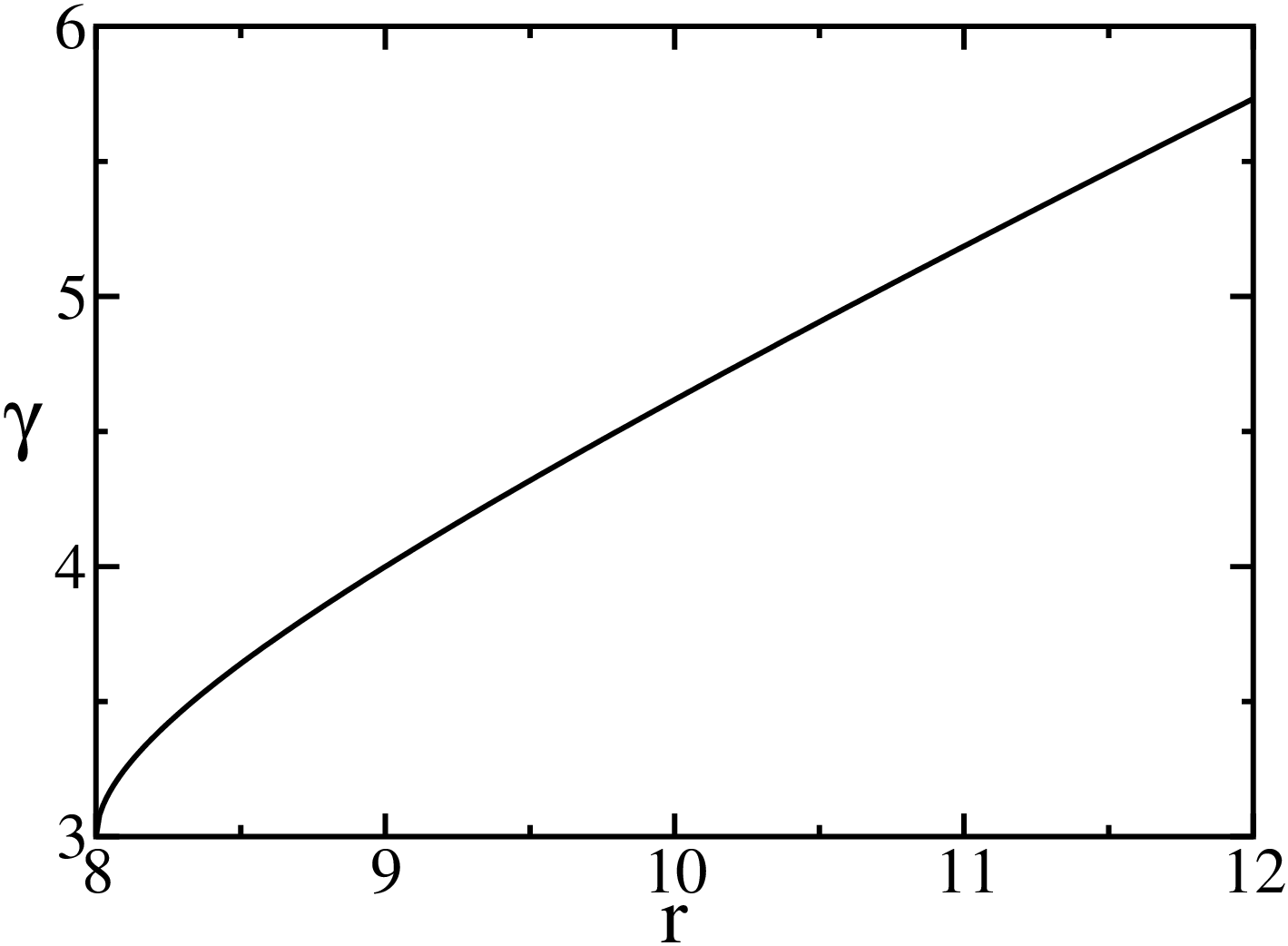}
\caption{The exponent $\gamma$ versus $r$ for the product rate.}
\label{fig-gamma-prod}
\end{figure}

We first discuss the cluster phase where $f(1)=1$, and consequently,
the analysis is a straightforward generalization of the above.  The
first derivative at $z=1$ again satisfies $R^{-1}[f'(1)]^2-f'(1)+1=0$.
At the critical point $R_c=4$ and therefore $r_c=8$. Solving this
quadratic equation, the first derivative is
\begin{equation}
\label{fprime-sol-prod}
f'(1)=\frac{r}{4}\left(1-\sqrt{1-\frac{8}{r}}\right).
\end{equation}
The first derivative is now the ratio between the second and the first
moments of the cluster size distribution, $f'(1)=\langle
k^2\rangle/\langle k\rangle$. The tail behavior follows from the
singular component of the generating function as in (\ref{fz-def-sum})
\begin{equation}
\label{fz-exp-prod}
f(z)=\underbrace{1+f'(1)(z-1)+\cdots}_{f_{\rm reg}(z)}+
\underbrace{B(1-z)^{\gamma-2}+\cdots}_{f_{\rm sing}(z)}.
\end{equation}
This again implies the power-law decay (\ref{ck-tail-sum}) for the
cluster size distribution.  We note that the unit shift in the
exponent is due to the different definition of the generating function
(\ref{fz-def-prod}). From the governing equation (\ref{fz-eq-sum}),
the exponent $\gamma$ satisfies $(\gamma-1)f'(1)=R$, and the decay
exponent is
\begin{equation}
\gamma=1+\frac{2}{1-\sqrt{1-\frac{8}{r}}}.
\end{equation}
Similar to (\ref{gamma-sol-sum}), the characteristic exponent grows
linearly with the injection rate, $\gamma\simeq \frac{r}{2}$, at large
injection rates (Figure \ref{fig-gamma-prod}).  The minimum value,
$\gamma=3$, is achieved at the phase transition point.  A more careful
analysis is again required at the phase transition point; we find (the
analysis is essentially the same as in the case of the sum rate,
Appendix~\ref{log}) that $c_k\sim k^{-3}(\ln k)^{-2}$ as $k\to\infty$.

\begin{figure}[t]
\includegraphics[width=0.45\textwidth]{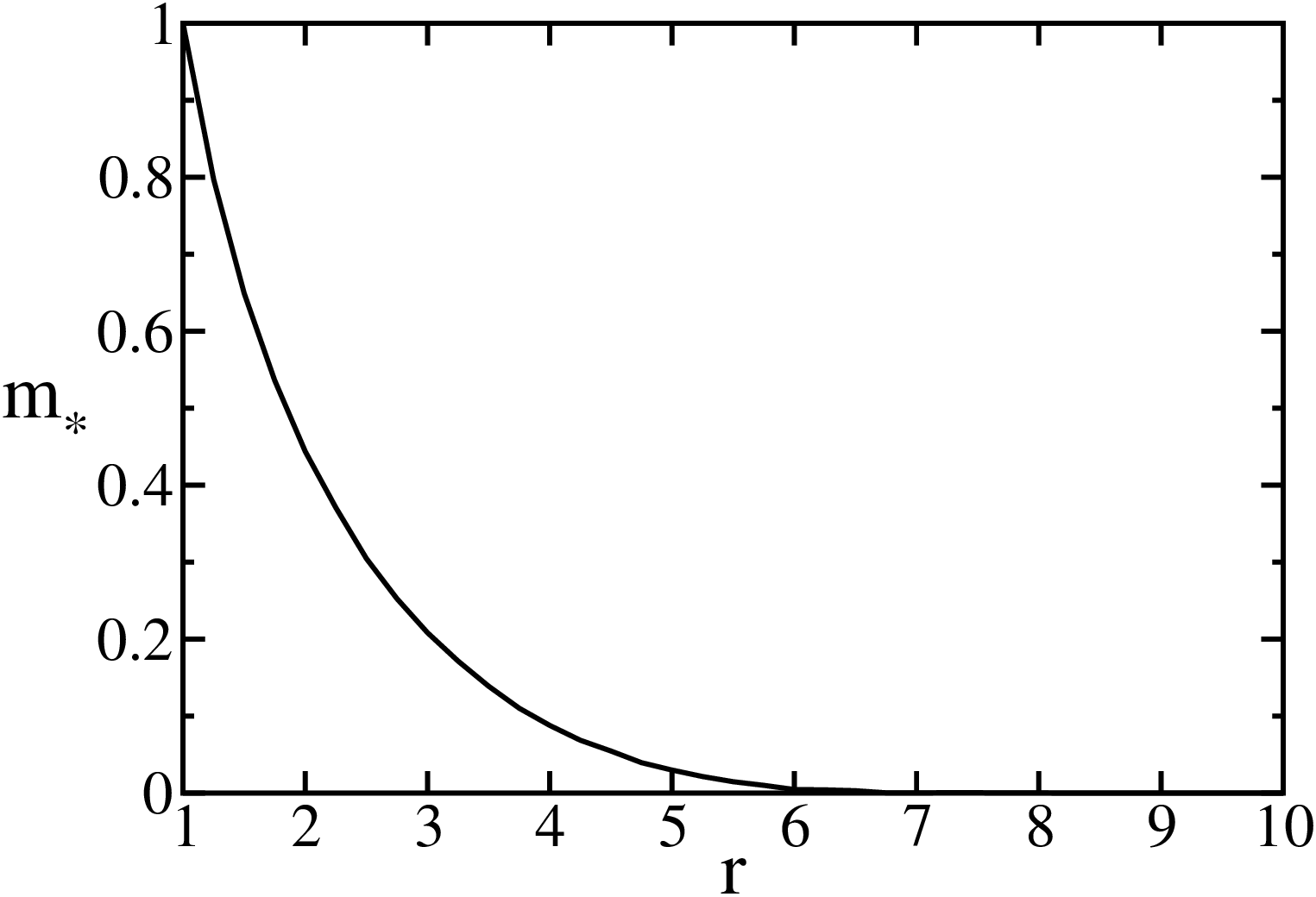}
\caption{The condensate mass $m_*$ as a function of $r$ obtained from
Monte Carlo simulations of the aggregation process.}
\label{fig-m*}
\end{figure}

In the condensate phase, we are able to obtain the mass of the
condensate only in the vicinity of the phase transition,
\begin{equation}
\label{m*-prod}
m_*\sim \exp\left[-\frac{\pi\sqrt{8}}{\sqrt{r-8}}\right],
\end{equation}
as $r\downarrow 8$.  The derivation of this result resembles that of
Refs.~\cite{kd,dms}; it is detailed in Appendix~\ref{giant}. In
contrast with the sum rate, the phase transition is now very gentle:
all derivatives of the condensate mass vanish at the transition point
$r=r_c$ \cite{vlt,kt}. In practice, it may be difficult to locate the
phase transition point (Figure \ref{fig-m*}).  We comment that similar
behavior was recently found in several models of growing networks
\cite{kd,chkns,dms,kr,kkkr,bb1,cb}. 

We performed Monte Carlo simulations to obtain the condensate mass
shown in Figure \ref{fig-m*}. In the simulations there are two
elemental steps: injection with probability $r/(r+1)$ and aggregation
with probability $1/(r+1)$. In an injection step, a cluster with unit
mass is added into the system. In an aggregation step, two clusters,
picked with probability proportional to the product of the two masses,
are merged. The simulations results represent a single run with a
total mass of $M=10^6$.

As a signature of the phase transition, the quantity $f'(1)$ has a
discontinuity at the phase transition point.  In the condensate phase,
this quantity is obtained, using the fact that $f(1)<1$, directly from
the governing equation (\ref{fz-eq-sum}), $f'(1)=R=r/2$. In the
cluster phase it is given by (\ref{fprime-sol-prod}). Therefore, the
ratio between the second and the first moments has a jump at $r_c=8$,
\begin{equation}
\label{jump}
\frac{\langle k^2\rangle}{\langle k\rangle}\to
\begin{cases}
4&r\uparrow r_c,\\
2&r\downarrow r_c.
\end{cases} 
\end{equation}
A similar jump occurs in the Berezinskii-Kosterlitz-Thouless phase
transition \cite{vlt,kt}.

Finally, we study the tail of the cluster size distribution. For this
purpose, it is convenient to make the transformation $z=e^w$ so that
the generating function (\ref{fz-def-prod}) is redefined
\begin{equation}
\label{fw-def}
f(w)=\sum_{k=1}^\infty k\,c_k e^{kw}.
\end{equation}
Substituting this definition into the recursion equation
(\ref{ck-eq-prod}), the generating function obeys
\begin{equation}
\label{fw-eq-prod}
\frac{df}{dw}=R\frac{f-e^w}{f-1}.
\end{equation}
The denominator suggests that $f(w)$ has a singularity at $w_0$ with
$f(w_0)=1$.  From the definition (\ref{fw-def}), this implies the
exponential decay $c_k\sim \exp(-k/k_0)$ with $k_0=1/w_0$. Since we
can not solve for the generating function, we can not locate this
singularity explicitly. Nevertheless, one can still deduce the
behavior near this singularity using asymptotic analysis.  Again, we
assume that the generating function has a regular component and a
singular component near $w=w_0$,
\begin{equation}
\label{fw-def-prod}
f(w)=\underbrace{1+\cdots}_{f_{\rm reg}(w)}+
\underbrace{A(w_0-w)^{\nu-2}+\cdots}_{f_{\rm sing}(w)}.
\end{equation}
Substituting this form into (\ref{fw-eq-prod}), and equating powers of
$(w_0-w)$ on both sides of the equation we obtain
\hbox{$\nu=5/2$}. Using the definition (\ref{fw-def}), the leading
behavior of the singular component $1-f(w_0)\sim (w_0-w)^{\nu-2}$
implies an algebraic correction to the leading exponential decay,
$c_k\sim k^{-\nu}\exp(-k/k_0)$ as $k\to \infty$.  We conclude that the
tail of the cluster size distribution decays as follows
\begin{equation}
\label{ck-tail-prod}
c_k\sim k^{-5/2}e^{-k/k_0}.
\end{equation}

The characteristic size can be related to the condensate mass in the
vicinity of the phase transition as the characteristic size is
expected to diverge in this limit, $k_0\to\infty$. Consequently, the
singularity is located close to the origin, $w_0\to 0$. Let us
estimate the behavior of the generating function near the origin. From
the definition (\ref{fw-def}), we have $f(w=0)=1-m_*$. Also, from the
governing equation we have $f'(0)=4$ in the limit $r\uparrow 8$ as in
(\ref{jump}). Therefore, $f(w)=1-m_*+4w$ as $w\to 0$ and from the
condition $f(w_0)=1$, the location of the singularity is
$w_0=m_*/4$. Therefore, the characteristic size grows as follows
\begin{equation}
k_0\sim 1/m_*
\end{equation}
in the vicinity of the phase transition point, $r\uparrow 8$. 

The tail behavior coincides with the critical behavior at sufficiently
small sizes, $c_k \sim k^{-3}$ for $k\ll k_0$, and exponential decay,
$c_k\sim \exp(-k/k_0)$, occurs beyond that scale, $k\gg k_0$. The two
behaviors should of course match at $k\approx k_0$: this implies the
proportionality constant in (\ref{ck-tail-prod}), $c_k\sim
k_0^{-1/2}k^{-5/2}\exp(-k/k_0)$.

\section{Conclusions}

In conclusion, we studied aggregation with constant injection of
mass. In this process, the total number of clusters grows with
time. For aggregation rates growing as the sum or the product of the
cluster sizes, there are two phases: a condensate phase and a cluster
phase.  In the condensate phase, a condensate containing a finite
fraction of the mass coexists with finite clusters, while in the
cluster phase there are only finite clusters.

For the sum rate, the mass of the condensate is a linear function of
the injection rate. Also, the cluster size distribution decays
algebraically in both phases and interestingly, the decay exponent is
a non-monotonic function of the injection rate. For the product rate
the condensate mass is extremely small in the vicinity of the phase
transition point and consequently, the phase transition is very
gentle. In this case, the tail of the cluster size distribution is
exponential in the condensate phase but algebraic in the cluster
phase. 

We comment that there are two frameworks for describing aggregation
processes: the Flory approach that allows interaction between giant
and finite clusters \cite{pjf} and the Stockmayer approach that allows
for interactions between finite clusters only \cite{whs}. We used the
more challenging former approach as it is the appropriate approach for
modeling growing random structures \cite{bk1}.

The various aggregation processes correspond to different random
growing structures, but our study focused only on the size of these
structures. We note that this theoretical framework can be generalized
to also study structural properties such as paths and cycles
\cite{bk}.

There are a number of possible extensions of this work. We focused on
the three classic aggregation rates where the generating function
obeys closed equations. This framework does not allow derivation of
the necessary conditions for the emergence of a condensate as a
function of the aggregation rate. Based on the sensitive algebraic
behavior in both of the phases, we speculate that the sum rate may be
the marginal case for condensation.

\noindent{\bf Acknowledgments.} We thank the Isaac Newton Institute
for Mathematical Sciences (Cambridge, England) and the Max Planck
Institute for Physics of Complex Systems (Dresden, Germany), where
this research was largely performed, for their hospitality. We
acknowledge financial support from DOE grant DE-AC52-06NA25396 and NSF
grant CHE-0532969.

%%%%%%%%%%%%%%%%%%%%%%%%%%%%%%%%%%%%%%%%%%%%
\appendix

\section{Derivation of (\ref{ns-crit})}
\label{log}

At the phase transition point $r=2$ we have $\gamma=2$ and therefore
the leading singular term \hbox{$f_{\rm sing}\propto
(1-z)^{\gamma-1}$} becomes regular. This suggests to use instead
\hbox{$f_{\rm sing}\propto (1-z)u(z)$} where $u(z)$ vanishes slower
than any power of $(1-z)$ as $z\to 1$. Thus at the phase transition
point we employ the following expansion
\begin{equation}
\label{exp-mod}
f(z)=1+2(z-1)+(z-1)u(z)+\ldots
\end{equation}
Substituting (\ref{exp-mod}) into (\ref{fz-eq-sum}) yields the 
differential equation
\begin{equation}
(z-1)\,\frac{du}{dz}+\frac{u^2}{2+u}=0
\end{equation}
whose (implicit) solution is
\begin{equation}
\label{uz}
-\frac{2}{u}+\ln u+\ln(1-z)={\rm const}.
\end{equation}
In the limit $z\to 1$, the integration constant is negligible compared
with the logarithmic term in the limit $z\to 1$ and consequently,
$u\to 2/\ln(1-z)$.  Indeed, $u$ vanishes slower than any power of
$(1-z)$ as $z\to 1$. Thus
\begin{equation}
\label{exp-crit}
f(z)=1+2(z-1)+\frac{2(z-1)}{\ln(1-z)}+\ldots
\end{equation}
Inverting this expansion leads to Eq.~(\ref{ns-crit}) \cite{kd}.

\section{Derivation of (\ref{m*-prod})}
\label{giant}

The mass of the condensate follows from the behavior of the generating
function at $z=1$, $m_*=1-f(1)$. To analyze the behavior near this
region, we make the transformations
\begin{subequations}
\begin{align}
\label{fz-trans}
f(z)&=1+x\,g(x),\\
\label{z-trans}
z&=1-x.
\end{align}
\end{subequations}
With these transformations, the equation for the generating function
(\ref{fz-eq-sum}) is transformed into the following first-order
nonlinear differential equation
\begin{equation}
\label{g-eq}
xgg'+g^2+Rg+R=0.
\end{equation}
In writing this equation, we kept only the leading order
terms. Writing
\hbox{$\frac{g}{g^2+Rg+R}dg+\frac{1}{x}dx$=0}, and integrating, 
we have
\begin{equation*}
\frac{1}{2}\ln (g^2+Rg+R)+\ln x
-\frac{R}{2a}\tan^{-1}\left(\frac{g+R/2}{a}\right)={\rm const}
\end{equation*}
where $a=\sqrt{R-R^2/4}$.  The integration constant can be evaluated
by taking the $x\to 0$ limit. Using \hbox{$m_*=-\lim_{x\to 0} xg(x)$},
the first two terms in the above equation approach $\ln m_*$ in the
limit $x\to 0$. Using \hbox{$\lim_{x\to 0}g(x)=-\infty$}, the last
term approaches $\frac{R\pi}{4a}$. Hence
\begin{eqnarray}
\frac{1}{2}\ln (g^2+Rg+R)+\ln x
&-&\frac{R}{2a}\tan^{-1}\left(\frac{g+R/2}{a}\right)\nonumber\\
&=&\ln m_*+\frac{\pi R}{4a}.
\label{integ}
\end{eqnarray}

Since we are interested in the behavior near the phase transition 
point, we take the limit $r\to 8$. In this limit, we can replace $R$
by $4$ and also, the quantity $g^2+Rg+R$ by $(g+2)^2$. Additionally,
we may replace $\tan^{-1}\frac{g+2}{a}$ by
$\frac{\pi}{2}-\frac{a}{g+2}$. With these substitutions,
Eq.~(\ref{integ}) becomes
\begin{equation}
\label{g-int-near}
\ln(-g-2)+\ln x+\frac{2}{g+2}=\ln m_*+\frac{2\pi}{a}.
\end{equation}

Next, we evaluate the left-hand side precisely at the phase transition
point, $r=8$. The critical behavior is detailed in Appendix
A. Substituting $g=-2-u$ and $x=1-z$ into (\ref{uz}), gives
\begin{equation}
\label{g-int-at}
\ln(-g-2)+\ln x+\frac{2}{g+2}={\rm const}.
\end{equation}
Substituting this into (\ref{g-int-near}) we obtain the condensate
mass in the vicinity of the phase transition (\ref{m*-prod}).

\end{document}